\font\bbb=msbm10                                                   

\def\C{\hbox{\bbb C}}

\def\AM{{\sl Advances in Math.}}

\def\MA{{\sl Math.\ Ann.}}

\def\PAMS{{\sl Proc.\ Amer.\ Math.\ Soc.}}

\def\PNAS{{\sl Proc.\ Nat.\ Acad.\ Sci.\ U.S.A.}}

\def\PRA{{\sl Phys.\ Rev.\ A}}

\def\PRL{{\sl Phys.\ Rev.\ Lett.}}
\def\PRSLA{{\sl Proc.\ Roy.\ Soc.\ Lond.\ A}}

\def\SIGACTN{{\sl SIGACT News}}

\def\vonneumann{\hbox{J. von Neumann}}

\def\hfb{\hfil\break}

\catcode`@=11
\newskip\ttglue

   \font\ninerm=cmr9    \font\eightrm=cmr8   \font\sixrm=cmr6
  \font\ninebf=cmbx9   \font\eightbf=cmbx8  \font\sixbf=cmbx6
  \font\nineit=cmti9   \font\eightit=cmti8  
  \font\ninesl=cmsl9   \font\eightsl=cmsl8  
  \font\ninemi=cmmi9   \font\eightmi=cmmi8  \font\sixmi=cmmi6

\font\bigtenbf=cmr10 scaled\magstep2 

\def\ninepoint{\def\rm{\fam0\ninerm}%
  \textfont0=\ninerm \scriptfont0=\sixrm
  \textfont1=\ninemi \scriptfont1=\sixmi
  \textfont\itfam=\nineit  \def\it{\fam\itfam\nineit}%
  \textfont\slfam=\ninesl  \def\sl{\fam\slfam\ninesl}%
  \textfont\bffam=\ninebf  \scriptfont\bffam=\sixbf
    \def\bf{\fam\bffam\ninebf}%
  \tt \ttglue=.5em plus.25em minus.15em
  \normalbaselineskip=11pt
  \setbox\strutbox=\hbox{\vrule height8pt depth3pt width0pt}%
  \normalbaselines\rm}

\def\eightpoint{\def\rm{\fam0\eightrm}%
  \textfont0=\eightrm \scriptfont0=\sixrm
  \textfont1=\eightmi \scriptfont1=\sixmi
  \textfont\itfam=\eightit  \def\it{\fam\itfam\eightit}%
  \textfont\slfam=\eightsl  \def\sl{\fam\slfam\eightsl}%
  \textfont\bffam=\eightbf  \scriptfont\bffam=\sixbf
    \def\bf{\fam\bffam\eightbf}%
  \tt \ttglue=.5em plus.25em minus.15em
  \normalbaselineskip=9pt
  \setbox\strutbox=\hbox{\vrule height7pt depth2pt width0pt}%
  \normalbaselines\rm}

\def\sfootnote#1{\edef\@sf{\spacefactor\the\spacefactor}#1\@sf
      \insert\footins\bgroup\eightpoint
      \interlinepenalty100 \let\par=\endgraf
        \leftskip=0pt \rightskip=0pt
        \splittopskip=10pt plus 1pt minus 1pt \floatingpenalty=20000
        \parskip=0pt\smallskip\item{#1}\bgroup\strut\aftergroup\@foot\let\next}
\skip\footins=12pt plus 2pt minus 2pt
\dimen\footins=30pc

\def\ie{{\it i.e.}}

\def\Theorem{T{\eightpoint HEOREM}}

\def\endproof{\vrule height6pt width6pt depth0pt}

\def\PQpf{PQ P{\eightpoint ENNY} F{\eightpoint LIP}}
\global\setbox3=\hbox{$\nearrow$}
\def\upxarrows{\nearrow\kern-\wd3\nwarrow}

\def\Simon{1}
\def\Shor{2}
\def\Qcrypto{3}
\def\Qcomm{4}
\def\vNM{5}
\def\vNquantum{6}
\def\Milnor{7}
\def\FarhiGutmann{8}
\def\Krauss{9}
\def\vNgame{10}
\def\Nashdef{11}
\def\Dirac{12}
\def\Grover{13}
\def\Glicksberg{14}
\def\Nashproof{15}
\def\cavityQED{16}
\def\iontrap{17}
\def\nmr{18}

\magnification=1200

\dimen0=\hsize \divide\dimen0 by 13 \dimendef\chasm=0
\dimen1=\hsize \advance\dimen1 by -\chasm \dimendef\usewidth=1
\dimen2=\usewidth \divide\dimen2 by 2 \dimendef\halfwidth=2
\dimen3=\usewidth \divide\dimen3 by 3 \dimendef\thirdwidth=3
\dimen4=\hsize \advance\dimen4 by -\halfwidth \dimendef\secondstart=4
\dimen5=\halfwidth \advance\dimen5 by -10pt \dimendef\indenthalfwidth=5
\dimen6=\thirdwidth \multiply\dimen6 by 2 \dimendef\twothirdswidth=6
\dimen7=\twothirdswidth \divide\dimen7 by 4 \dimendef\qttw=7
\dimen8=\qttw \divide\dimen8 by 4 \dimendef\qqttw=8
\dimen9=\qqttw \divide\dimen9 by 4 \dimendef\qqqttw=9

\parskip=0pt\parindent=0pt

\line{\hfil February 1998}
\line{\hfil quant-ph/9804010}
\vfill
\centerline{\bf\bigtenbf QUANTUM STRATEGIES}
\bigskip\bigskip
\centerline{\bf David A. Meyer}
\bigskip 
\centerline{\sl Center for Social Computation,
                Institute for Physical Sciences, and}
\smallskip
\centerline{\sl Project in Geometry and Physics}
\centerline{\sl Department of Mathematics}
\centerline{\sl University of California/San Diego}
\centerline{\sl La Jolla, CA 92093-0112}
\centerline{dmeyer@chonji.ucsd.edu}
\vfill
\centerline{ABSTRACT}
\bigskip
\noindent We consider game theory from the perspective of quantum
algorithms.  Strategies in classical game theory are either pure
(deterministic) or mixed (probabilistic).  We introduce these basic 
ideas in the context of a simple example, closely related to the 
traditional M{\eightpoint ATCHING} P{\eightpoint ENNIES} game.  While 
not every two-person zero-sum finite game has an equilibrium in the 
set of pure strategies, von~Neumann showed that there is always an 
equilibrium at which each player follows a mixed strategy.  A mixed 
strategy deviating from the equilibrium strategy cannot increase a 
player's expected payoff.  We show, however, that in our example a 
player who implements a quantum strategy {\sl can\/} increase his 
expected payoff, and explain the relation to efficient quantum 
algorithms.  We prove that in general a quantum strategy is always at 
least as good as a classical one, and furthermore that when both 
players use quantum strategies there need not be any equilibrium, but 
if both are allowed mixed quantum strategies there must be.

\bigskip\bigskip
\noindent 1996 Physics and Astronomy Classification Scheme:
                   03.65.-w, 
                   89.80.+h. 

\noindent American Mathematical Society Subject Classification:
                   81P15,    
                   90D05.    

\noindent{\sl Journal of Economic Literature\/} Classification System:
                   C72.      

\global\setbox1=\hbox{Key Words:\enspace}
\parindent=\wd1
\item{Key Words:}  quantum computation; game theory. 

\vfill
\eject

\headline{\ninepoint\it Quantum strategies       \hfil David A. Meyer}
\parskip=10pt
\parindent=20pt

Attention to the physical representation of information underlies the 
recent theories of quantum computation, quantum cryptography and 
quantum communication.  In each case representation in a quantum 
system provides advantages over the classical situation:  Simon's 
quantum algorithm to identify the period of a function chosen by an 
oracle is more efficient than any deterministic or probabilistic 
algorithm [\Simon] and provided the foundation for Shor's polynomial
time quantum algorithm for factoring [\Shor].  The quantum protocols 
for key distribution devised by Wiener, Bennett and Brassard, and 
Ekert are qualitatively more secure against eavesdropping than any 
classical cryptosystem [\Qcrypto].  And Cleve and Buhrman, and 
van~Dam, H{\o}yer and Tapp have shown that prior quantum entanglement
reduces communication complexity [\Qcomm].  In this report we add
game theory to the list:  quantum strategies can be more successful
than classical ones.

While this result may seem obscure or surprising, in fact it is 
neither.  Cryptographic situations, for example, are readily conceived
as games; it is reasonable to ask if the advantages of quantum key
distribution generalize.  Game theory, on the other hand, seems to beg
for a quantum version:  Classical strategies can be pure or mixed; the 
correspondence of this nomenclature, due to von~Neumann [\vNM], with 
that of quantum mechanics is surely no accident [\vNquantum].  
Furthermore, games against nature, originally studied by Milnor 
[\Milnor], should include those for which nature is quantum 
mechanical.  Finally, in their extensive form, games are represented 
by `trees' [\vNM], just as are (quantum) algorithms 
[\Simon,\FarhiGutmann].  We will exploit this similarity to analyze 
the effectiveness of quantum strategies, exemplified in the following 
very simple game:

\noindent\PQpf:  The starship {\sl Enterprise\/} is facing some 
immanent---and apparently inescapable---calamity when Q appears on the 
bridge and offers to help, provided Captain Picard%
\sfootnote{$^*$}{Captain Picard and Q are characters in the popular 
American television (and movie) series {\sl Star Trek:  The Next 
Generation\/} whose initials and abilities are ideal for this
illustration.  See [\Krauss].}
can beat him at penny flipping:  Picard is to place a penny head up in 
a box, whereupon they will take turns (Q, then Picard, then Q) 
flipping the penny (or not), without being able to see it.  Q wins if 
the penny is head up when they open the box.

This is a two-person zero-sum strategic game which might be analyzed 
traditionally using the payoff matrix:
$$
\vbox{\offinterlineskip
\halign{&\vrule#&\strut\enspace\hfil#\enspace\cr
\omit&\omit&\omit&$NN$&\omit&$NF$&\omit&$FN$&\omit&$FF$&\omit\cr
\omit&\omit&\multispan9\hrulefill\cr
\omit&\omit&height2pt&\omit&&\omit&&\omit&&\omit&\cr
\omit&$N$&&$-1$&&1&&1&&$-1$&\cr
\omit&\omit&height2pt&\omit&&\omit&&\omit&&\omit&\cr
\omit&\omit&\multispan9\hrulefill\cr
\omit&\omit&height2pt&\omit&&\omit&&\omit&&\omit&\cr
\omit&$F$&&1&&$-1$&&$-1$&&1&\cr
\omit&\omit&height2pt&\omit&&\omit&&\omit&&\omit&\cr
\omit&\omit&\multispan9\hrulefill\cr
}}
$$
where the rows and columns are labelled by Picard's and Q's {\sl pure
strategies}, respectively; $F$ denotes a flip and $N$ denotes no flip; 
and the numbers in the matrix are Picard's payoffs:  1 indicating a 
win and $-1$ a loss.%
\sfootnote{$^{\dagger}$}{Since when one player wins, the other loses, 
we need only list one player's payoffs; whenever this is the case the 
game is called {\sl zero-sum}.  {\sl Strategic\/} refers to the fact 
that the players choose their strategies independently of the other 
player's actions [\vNgame,\vNM].}
For example, consider the top entry in the second column:  Q's 
strategy is to flip the penny on his first turn and then not flip it 
on his second, while Picard's strategy is to not flip the penny on his 
turn.  The result is that the state of the penny is, successively:  
$H$, $T$, $T$, $T$, so Picard wins.

Having studied game theory in his Advanced Decision Making course at
Starfleet Academy, Captain Picard has no difficulty determining his
optimal strategy:  Suppose he doesn't flip the penny.  Then if Q flips
it an even number of times, Picard loses.  Similarly, if Picard flips
the penny, then if Q flips it only once, Picard loses.  Thus \PQpf\ 
has no deterministic solution [\vNM], no deterministic Nash 
equilibrium [\Nashdef]:  there is no pair of pure strategies, one for 
each player, such that neither player can improve his result by 
changing his strategy while the other player does not.  But, as 
von~Neumann proved there must be [\vNgame,\vNM], since this is a 
two-person zero-sum strategic game with only a finite number of 
strategies, there is a probabilistic solution:  It is easy to check 
that the pair of {\sl mixed\/} strategies consisting of Picard 
flipping the penny with probability ${1\over2}$ and Q playing each of 
his four strategies with probability ${1\over4}$ is a 
{\sl probabilistic\/} Nash equilibrium:  neither player can improve 
his {\sl expected\/} payoff (which is 0 in this case) by changing the 
probabilities with which he plays each of his pure strategies while 
the other player does not.

Figuring his chances of winning are $1/2$, Captain Picard agrees to 
play.  But he loses.  The rules of the game allow Q two moves so, 
although his analysis indicates no benefit for Q from the second move,
Picard tries arguing that they should therefore play several times.
To his surprise Q agrees---and proceeds to beat Picard the next 9 
times as well.  Picard is sure that Q is cheating.  Is he?

\moveright\secondstart\vtop to 0pt{\hsize=\halfwidth
$$
\matrix{    H    &            &     T    \cr
        \uparrow & \upxarrows & \uparrow \cr
            H    &            &     T    \cr
        \uparrow & \upxarrows & \uparrow \cr
            H    &            &     T    \cr
        \uparrow &  \nearrow  &          \cr
            H    &            &          \cr
       }
$$
\vskip 0.25\baselineskip
\eightpoint{%
\noindent{\bf Figure~1}.  PQ P{\sixrm ENNY} F{\sixrm LIP} in
extensive form.
}}
\vskip -\baselineskip
\parshape=11
0pt \halfwidth
0pt \halfwidth
0pt \halfwidth
0pt \halfwidth
0pt \halfwidth
0pt \halfwidth
0pt \halfwidth
0pt \halfwidth
0pt \halfwidth
0pt \halfwidth
0pt \hsize
To understand what Q is doing, let us reanalyze \PQpf\ in terms of the
sequence of moves---in its {\sl extensive form}.  Conventionally the
extensive form of a game is illustrated by a tree with a distinct 
vertex for each partial sequence of player actions and outgoing edges
from each vertex corresponding to the possible actions on the next 
move.  For our purposes it is more useful to study the quotient of
this tree obtained by identifying the vertices at which both the state
of the game and the number of preceding moves are the same.  Thus we
illustrate the extensive form of \PQpf, not with a binary tree of 
height 3, but with the directed graph shown in Figure~1.  The vertices
are labelled $H$ or $T$ according to the state of the penny and each 
diagonal arrow represents a flip while each vertical arrow represents
no flip.

Now it is natural to define a two dimensional vector space $V$ with
basis $\{H,T\}$ and to represent player strategies by sequences of 
$2\times 2$ matrices.  That is, the matrices
$$
F := \bordermatrix{     
                       &\scriptstyle{H} & \scriptstyle{T}          \cr
       \scriptstyle{H} &       0        &        1                 \cr
       \scriptstyle{T} &       1        &        0                 \cr
                  }
\qquad\hbox{and}\qquad
N := \bordermatrix{     
                       &\scriptstyle{H} & \scriptstyle{T}          \cr
       \scriptstyle{H} &       1        &        0                 \cr
       \scriptstyle{T} &       0        &        1                 \cr
                  }
$$
correspond to flipping and not flipping the penny, respectively,
since we define them to act by left multiplication on the vector
representing the state of the penny.  A {\sl mixed\/} action is a 
convex linear combination of $F$ and $N$, which acts as a $2\times 2$ 
(doubly) stochastic matrix:
$$
\bordermatrix{     
                       &\scriptstyle{H} & \scriptstyle{T}          \cr
       \scriptstyle{H} &      1-p       &        p                 \cr
       \scriptstyle{T} &       p        &       1-p                \cr
                  }
$$
if the player flips the penny with probability $p \in [0,1]$.  A
sequence of mixed actions puts the state of the penny into a convex
linear combination $a H + (1-a) T$, $0 \le a \le 1$, which means that 
if the box is opened the penny will be head up with probability $a$.

Q, however, is eponymously utilizing a {\sl quantum\/} strategy, 
namely a sequence of unitary, rather than stochastic, matrices.  In 
standard Dirac notation [\Dirac] the basis of $V$ is written 
$\{|H\rangle,|T\rangle\}$.  A {\sl pure\/} quantum state for the penny 
is a linear combination $a|H\rangle + b|T\rangle$, $a,b \in \C$, 
$a\overline{a} + b\overline{b} = 1$, which means that if the box is 
opened, the penny will be head up with probability $a\overline{a}$.  
Since the penny starts in state $|H\rangle$, this is the state of the 
penny if Q's first action is the unitary operation
$$
U_1 = U(a,b) :=
\bordermatrix{
                       &\scriptstyle{H} & \scriptstyle{T}          \cr
       \scriptstyle{H} &       a        &        b                 \cr
       \scriptstyle{T} &  \overline{b}  &  -\overline{a}           \cr
             }.
$$

Recall that Captain Picard is also living up to his initials, 
utilizing a {\sl c\/}lassical {\sl p\/}robabil\-istic strategy in 
which he flips the penny with probability $p$.  After his action the 
penny is in a {\sl mixed\/} quantum state, \ie, it is in the pure 
state $b|H\rangle + a|T\rangle$ with probability $p$ and in the pure 
state $a|H\rangle + b|T\rangle$ with probability $1-p$.  Mixed states 
are conveniently represented as {\sl density matrices\/} [\vNquantum], 
elements of $V \otimes V^{\dagger}$ with trace 1; the diagonal entry 
$(i,i)$ is the probability that the system is observed to be in state 
$|i\rangle$.  The density matrix for a pure state $|\psi\rangle \in V$ 
is the projection matrix $|\psi\rangle\langle\psi|$ and the density 
matrix for a mixed state is the corresponding convex linear 
combination of pure density matrices.  Unitary transformations act on
density matrices by conjugation:  The penny starts in the pure state
$\rho_0 = |H\rangle \langle H|$ and Q's first action puts it into the
pure state:
$$
\rho_1 = U_1^{\vphantom\dagger} \rho_0 U_1^{\dagger} 
       = \pmatrix{ a\overline{a} & a\overline{b} \cr
                   b\overline{a} & b\overline{b} \cr
                 }.
$$
Picard's mixed action acts on this density matrix, not as a stochastic
matrix on a probabilistic state, but as a convex linear combination of
unitary (deterministic) transformations:
$$
\rho_2 = p F \rho_1 F^{\dagger} + (1-p) N \rho_1 N^{\dagger}
       = \pmatrix{ pb\overline{b} + (1-p)a\overline{a} &
                   pb\overline{a} + (1-p)a\overline{b} \cr
                   pa\overline{b} + (1-p)b\overline{a} &
                   pa\overline{a} + (1-p)b\overline{b} \cr
                 }.                                           \eqno(1)
$$
For $p = {1\over2}$ the diagonal elements of $\rho_2$ are each 
${1\over2}$.  If the game were to end here, Picard's strategy would 
ensure him an expected payoff of 0, independently of Q's strategy.  In 
fact, if Q were to employ any strategy for which 
$a\overline{a} \ne b\overline{b}$, Picard could obtain an expected 
payoff of $|a\overline{a} - b\overline{b}| > 0$ by setting $p = 0,1$ 
according to whether $b\overline{b} > a\overline{a}$, or the reverse.
Similarly, if Picard were to choose $p \ne {1\over2}$, Q could obtain 
an expected payoff of $|2p - 1|$ by setting $a = 1$ or $b = 1$ 
according to whether $p < {1\over2}$, or the reverse.  Thus the 
mixed/quantum equilibria for the two-move game are pairs 
$\bigl([{1\over2}F + {1\over2}N],[U(a,b)]\bigr)$ for which 
$a\overline{a} = {1\over2} = b\overline{b}$ and the outcome is the 
same as if both players utilize optimal mixed strategies.

But Q has another move $U_3$ which again transforms the state of the 
penny by conjugation to 
$\rho_3 = U_3^{\vphantom\dagger}\rho_2 U_3^{\dagger}$.  If Q's 
strategy consists of $U_1 = U(1/\sqrt{2},1/\sqrt{2}) = U_3$, his first
action puts the penny into a simultaneous eigenvalue 1 eigenstate of
both $F$ and $N$, which is therefore invariant under {\sl any\/} mixed 
strategy $pF + (1-p)N$ of Picard; and his second action inverts his
first to give $\rho_3 = |H\rangle\langle H|$.  That is, with 
probability 1 the penny is head up!  Since Q can do no better than to 
win with probability 1, this is an optimal quantum strategy for him.  
All the pairs 
$\bigl([pF + (1-p)N],
       [U(1/\sqrt{2},1/\sqrt{2}),U(1/\sqrt{2},1/\sqrt{2})]\bigr)$ 
are mixed/quantum equilibria for \PQpf, with value $-1$ to Picard; 
this is why he loses every game.

\PQpf\ is a very simple game, but it is structurally similar to the 
oracle problems for which efficient quantum algorithms are 
known---with Picard playing the role of the oracle.  In Simon's 
problem the functions $f : \{0,1\}^n \to \{0,1\}^n$ which satisfy
$f(x) = f(y)$ if and only if $y = x \oplus s$ for some 
$s \in \{0,1\}^n$ ($\oplus$ denotes componentwise addition, mod 2), 
correspond to Picard's pure strategies; we may imagine the oracle 
choosing a mixed strategy intended to minimize our chances of 
efficiently determining $s$ probabilistically.  Simon's algorithm is a 
quantum strategy which is more successful than any mixed, \ie, 
probabilistic, one [\Simon].  Similarly, in the problem of searching a 
database of size $N$, the locations in the database correspond to pure 
strategies; again we may imagine the oracle choosing a mixed strategy 
designed to frustrate our search for an item at some specified 
location.  Grover's algorithm is a quantum strategy for a game of $2m$ 
moves alternating between us and the oracle, where $m = O(\sqrt{N})$, 
which out performs any mixed strategy [\Grover].  These three examples 
suggest the following:

\noindent\Theorem\ 1:  {\sl There is always a mixed/quantum 
equilibrium for a two-person zero-sum game, at which the expected
payoff for the player utilizing a quantum strategy is at least as
great as his expected payoff with an optimal mixed strategy.}

\noindent{\sl Proof\/} (sketch):  A sequence of mixed actions puts the 
game into a convex linear combination $\sum p_i|i\rangle$ of pure
states.  If one of the players utilizes a quantum strategy, the state
of the game is described instead by a density matrix.  We must show
that there is always a quantum strategy which reproduces the $p_i$ as 
the diagonal elements in the density matrix.  Assume by induction that
this is true up to a move of the classical player.  His action has the
same effect on the diagonal elements of the density matrix as it does
on the $p_i$ in the original mixed/mixed equilibrium move sequence.
(See (1).)  All that remains to be shown is that a single action of 
the quantum player can be chosen to reproduce the effect of a mixed
action.  It is only necessary to consider $U(2)$ actions on a general
$2 \times 2$ density matrix.  If the phase of the $(1,2)$ element in
the density matrix is $\gamma$, a straightforward calculation verifies
that the unitary matrix 
$U(e^{\pi/2 - \gamma}\sqrt{\vphantom{1}p},\sqrt{1-p})$ reproduces the 
effect of the mixed action $pF + (1-p)N$ on the diagonal elements.
                                                       \hfill\endproof

Of course, the more interesting question is for which games there is a
quantum strategy which improves upon the optimal mixed strategy.  By
the analogy with algorithms, this is essentially the fundamental 
question of which problems can be solved more efficiently by quantum 
algorithms than by classical ones.  We may hope that the game 
theoretic perspective will suggest new possibilities for efficient 
quantum algorithms.

Another natural question to ask is what happens if both players 
utilize quantum strategies.  By considering \PQpf\ we can prove the 
following:

\noindent\Theorem\ 2:  {\sl A two-person zero-sum game need not have a 
quantum/quantum equilibrium.}

\noindent{\sl Proof\/}:  Consider an arbitrary pair of quantum
strategies $([U_2],[U_1,U_3])$ for \PQpf.  Suppose 
$U_3 U_2 U_1 |H\rangle \ne |H\rangle$.  Then Q can improve his 
expected payoff (to 1) by changing his strategy, replacing $U_3$ with
$U_1^{-1} U_2^{-1}$, which is unitary since $U_1$ and $U_2$ are.  
Similarly, suppose $U_3 U_2 U_1 |H\rangle \ne |T\rangle$.  Then 
Picard can improve his expected payoff (to 1) by changing his 
strategy, replacing $U_2$ with $U_3^{-1} F U_1^{-1}$, which is unitary
since each of $U_1$, $U_3$ and $F$ is.  Since $U_3 U_2 U_1 |H\rangle$
cannot be both $|H\rangle$ and $|T\rangle$, at least one of the 
players can improve his expected payoff by changing his strategy while
the other does not.  Thus $([U_2],[U_1,U_3])$ cannot be an 
equilibrium, for any $U_1$, $U_2$, $U_3$, so \PQpf\ has no 
quantum/quantum equilibrium.                           \hfill\endproof

That is, the situation when both players utilize quantum strategies is 
the same as when they both utilize pure (classical) strategies:  there
need not be any equilibrium solution.  This suggests looking for the
analogue of von~Neumann's result on the existence of mixed strategy 
equilibria [\vNgame,\vNM].  So we should consider strategies which are 
convex linear combinations of unitary actions---{\sl mixed quantum\/} 
strategies.

\noindent\Theorem\ 3:  {\sl A two-person zero-sum game always has a
mixed quantum/mixed quantum equilibrium.}

\noindent{\sl Proof\/}:  Since mixed quantum actions form a convex 
compact subset of a finite dimensional vector space, this is an 
immediate corollary of Glicksberg's generalization [\Glicksberg] of 
Nash's proof [\Nashproof] for the existence of game equilibria. 
                                                       \hfill\endproof

Finally, we remark that while decoherence precludes the play of \PQpf\
with a penny, only a two state quantum system is logically necessary.   
Thus each of the physical systems in which quantum gate operations 
have been demonstrated---QED cavities [\cavityQED], ion traps 
[\iontrap] and NMR machines [\nmr]---could be used to realize games of 
\PQpf\ and even slightly more complicated games with quantum 
strategies.

\medskip
\noindent{\bf Acknowledgements}
\nobreak

\nobreak
\noindent I thank Ian Agol, Thad Brown, Mike Freedman, Jeong Han Kim,
Jeff Remmel and Francis Zane for discussions about this work.

\medskip
\global\setbox1=\hbox{[00]\enspace}
\parindent=\wd1

\noindent{\bf References}
\bigskip

\parskip=0pt
\item{[\Simon]}
D. R. Simon,
``On the power of quantum computation'',
in S. Goldwasser, ed.,
{\sl Proceedings of the 35th Symposium on Foundations of Computer 
Science}, Santa Fe, NM, 20--22 November 1994
(Los Alamitos, CA:  IEEE Computer Society Press 1994) 116--123.

\item{[\Shor]}
P. W. Shor,
``Algorithms for quantum computation:  discrete logarithms and 
  factoring'',
in S. Goldwasser, ed.,
{\sl Proceedings of the 35th Symposium on Foundations of Computer 
Science}, Santa Fe, NM, 20--22 November 1994
(Los Alamitos, CA:  IEEE Computer Society Press 1994) 124--134.

\item{[\Qcrypto]}
S. Wiesner, 
``Conjugate coding'',
\SIGACTN\ {\bf 15} (1983) 78--88;\hfb
C. H. Bennett and G. Brassard,
``Quantum cryptography:  Public-key distribution and coin tossing'',
in 
{\sl Proceedings of the IEEE International Conference on Computers,
Systems and Signal Processing}, Bangalore, India, December 1984
(New York:  IEEE 1984) 175--179;\hfb
A. Ekert,
``Quantum cryptography based on Bell's theorem'',
\PRL\ {\bf 67} (1991) 661--663.

\item{[\Qcomm]}
R. Cleve and H. Buhrman,
``Substituting quantum entanglement for communication'',
\PRA\ {\bf 56} (1997) 1201--1204;\hfb
W. van~Dam, P. H{\o}yer and A. Tapp,
``Multiparty quantum communication complexity'',
preprint (1997), quant-ph/9710054.

\item{[\vNM]}
\vonneumann\ and O. Morgenstern,
{\sl Theory of Games and Economic Behavior}, third edition
(Princeton:  Princeton University Press 1953).

\item{[\vNquantum]}
\vonneumann,
{\it Mathematische Grundlagen der Quantenmechanik\/}
(Berlin:  Spring\-er-Verlag 1932); 
transl.\ by R. T. Beyer as
{\sl Mathematical Foundations of Quantum Mechanics\/}
(Princeton:  Princeton University Press 1955).

\item{[\Milnor]}
J. Milnor,
``Games against nature'',
in R. M. Thrall, C. H. Coombs and R. L. Davis, eds.,
{\sl Decision Processes\/}
(New York:  John Wiley \& Sons 1954) 49--59.

\item{[\FarhiGutmann]}
E. Farhi and S. Gutmann,
``Quantum computation and decision trees'',
MIT preprint CTP-2651 (1997), quant-ph/9706062.

\item{[\Krauss]}
L. M. Krauss,
{\sl The Physics of Star Trek}, 
with a forword by Stephen Hawking
(New York:  HarperCollins 1995).

\item{[\vNgame]}
\vonneumann,
``{\it Zur theorie der gesellschaftsspiele\/}'',
\MA\ {\bf 100} (1928) 295--320.

\item{[\Nashdef]}
J. F. Nash,
``Equilibrium points in $N$-person games'',
\PNAS\ {\bf 36} (1950) 48--49.

\item{[\Dirac]}
P. A. M. Dirac,
{\sl The Principles of Quantum Mechanics}, fourth edition
(Oxford:  Oxford University Press 1958).

\item{[\Grover]}
L. K. Grover,
``A fast quantum mechanical algorithm for database search'',
in 
{\sl Proceedings of the 28th Annual ACM Symposium on the Theory of
Computing}, Philadelphia, PA, 22--24 May 1996
(New York:  ACM 1996) 212--219.

\item{[\Glicksberg]}
I. L. Glicksberg,
``A further generalization of the Kakutani fixed point theorem, 
  with application to Nash equilibrium points'',
\PAMS\ {\bf 3} (1952) 170--174.

\item{[\Nashproof]}
J. F. Nash,
``Non-cooperative games'',
\AM\ {\bf 54} (1951) 286--295.

\item{[\cavityQED]}
Q. A. Turchette, C. J. Hood. W. Lange, H. Mabuchi and H. J. Kimble,
``Measurement of conditional phase shifts for quantum logic'',
\PRL\ {\bf 75} (1995) 4710--4713.

\item{[\iontrap]}
J. I. Cirac and P. Zoller,
``Quantum computation with cold trapped ions'',
\PRL\ {\bf 74} (1995) 4091--4094;\hfb
C. Monroe, D. M. Meekhof, B. E. King, W. M. Itano and 
  D. J. Wineland,
``Demonstration of a fundamental logic gate'',
\PRL\ {\bf 75} (1995) 4714--4717.

\item{[\nmr]}
D. G. Cory, M. D. Price, A. F. Fahmy and T. F. Havel,
``Nuclear magnetic resonance spectroscopy:  an experimentally 
  accessible paradigm for quantum computing'',
quant-ph/9709001;\hfb
I. L. Chuang, N. Gershenfeld, M. G. Kubinec and D. W. Leung,
``Bulk quantum computation with nuclear magnetic resonance:
  theory and experiment'',
\PRSLA\ {\bf 454} (1998) 447--467.

\bye